\documentclass[twocolumn,groupedaddress]{revtex4}

\usepackage[latin1]{inputenc}
\usepackage{graphicx}
\usepackage{amsmath}
\usepackage{amssymb}

\begin{document}

\title{Excess dissipation in a single-electron box: The Sisyphus
  resistance}

\author{F. Persson}
\email[]{fredrik.persson@chalmers.se}
\author{C.M. Wilson}
\author{M. Sandberg}
\author{G. Johansson}
\author{P. Delsing}
\email[]{per.delsing@chalmers.se}
\affiliation{Department of Microtechnology and Nanoscience (MC2),
Chalmers University of Technology, SE-412~96 G\"{o}teborg, Sweden}

\date{\today}


\begin{abstract}
We present measurements of the ac response of a single-electron
box (SEB). We apply an rf signal with a frequency larger than the
tunneling rate and drive the system out of equilibrium. We observe
much more dissipation in the SEB then one would expect from a simple
circuit model. We can explain this in terms of a mechanism that we
call the Sisyphus resistance. The Sisyphus resistance has a strong
gate dependence which can be used for electrometery applications.
\end{abstract}

\maketitle

Dissipation in quantum systems has been an important area of research
for many years. It has lately gained renewed attention as quantum
systems have increasingly come to the forefront of technology,
including research in nanotechnology and quantum information.  In many
of these contexts, dissipation can be modeled using a simple two-level
system (TLS) or an ensemble of them.  It is therefore generally useful
to study dissipation in TLSs. For instance, in the context of quantum
information \cite{ClarkeWilhelm}, where any dissipation causes unwanted
decoherence, TLSs are important in several different ways. First, the
basic building block of all quantum bits is an effective TLS. In
addition, the presence of parasitic TLSs in dielectrics has been shown
to result in losses which degrade the quality factor of electrical
resonators and limit the coherence of superconducting qubits
\cite{Martinis:DielectricLoss}. Fast driving of a TLS through an
avoided level crossings (ALC) recently received considerable
attention, and has been analyzed in terms of Landau-Zener (LZ)
transitions \cite{Shytov:LZ,Oliver,Sillanpaa} and dressed states
\cite{Wilson:LDS}. Recently, in a setup related to the one studied
here, Sisyphus cooling \cite{Wineland} and amplification was observed
in a superconducting circuit \cite{Grajcar:SCA}.

In this Letter, we study a mesoscopic circuit consisting of a small
metallic island connected by a tunnel junction to a much larger
reservoir. Charging effects (Coulomb blockade) result in well
defined energy levels which become degenerate at a specific bias
point. An ac drive is used to cyclically drive the system through this
level crossing. Due to the low transparency of the tunnel barrier, the
coupling between the levels is negligible and the probability of a LZ
transition when crossing the degeneracy point is very close to
unity. However, due to the large degeneracy of the electronic states
on the island, the total system can have a significant relaxation (or
tunneling) rate.
If the frequency of the drive is comparable to the relaxation
rate of the system, alternating excitation and relaxation of the system
lead to excess dissipation which can be directly measured. We call
this process the Sisyphus resistance. We develop a quantitative model
of the behavior that shows very good agreement with the measured
response. 

\begin{figure}[tb!]
 \includegraphics[width=0.9\columnwidth]{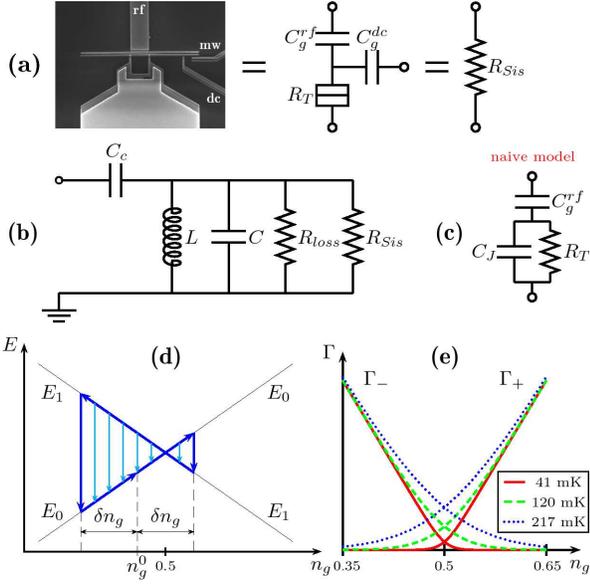}
 \caption{\label{fig:intro}
   (a) A scanning-electron micrograph of the SEB. The island of the
   SEB is 8 $\mu$m long and $100$ nm wide. The SEB has three
   different gates: one rf gate connecting the SEB to a parallel
   resonator used for readout, one dc gate to set the working point
   and one microwave gate used in the characterization. To the right
   of the micrograph are different circuit models used in the
   discussion of the measured response.
   (b) The circuit model used in analyzing the response. The SEB is
   modeled by the voltage dependent resistor $R_{Sis}$. The internal
   losses in the resonator, mostly due to the  normal top layer, are
   modeled as a parallel resistor $R_{loss}$. The coupling capacitor
   $C_c$ is used to decouple the resonator from the 50 $\Omega$
   impedance of the environment and thereby increase the quality
   factor of the resonator.
   (c) A naive model of the SEB, which is compared with the Sisyphus
   resistance in the text. $C_J$ is the geometric capacitance of the
   tunnel junction.
   (d) The process behind the Sisyphus resistance. $E_0$ and $E_1$
   are the electrostatic energies of the two charge states. Starting
   from a charge state of lower energy, the rf signal $\delta n_g$
   drives the system past the degeneracy point and into an excited
   state, at the same time charging the system. When the system
   eventually relaxes by tunneling, the charging energy is lost.
   (e) The tunnel rates between the two charge states as a function
   of the normalized gate voltage, $n_g$, for three different
   temperatures.}
\end{figure}

The charging effects which give rise to the quantization of the energy
levels have a very peculiar effect on dissipation in the circuit. Far
away from the degeneracy point, the Coulomb blockade prevents
tunneling, and results in a high Sisyphus resistance and low
dissipation. However, at the degeneracy point, the charging effects
result in a low Sisyphus resistance with dissipation much larger than
that expected from a gate capacitance in series with an Ohmic resistor
with the value of the tunneling resistance, Fig.~\ref{fig:intro}c.
We note that the Sisyphus mechanism described here is not limited to
Coulomb blockade devices, but should be a generic property of any
system with an energy level crossing. In an analogous way to the
Sisyphus mechanism discussed here, it was suggested that phonon
pumping of TLSs can dominate losses in micro-mechanical
resonators at low temperatures \cite{QuantumFriction} and one could
imagine having the same effect in electrical resonators. It has been
suggested that TLSs may dominate the loss in electrical resonators
under conditions important for quantum information applications
\cite{Martinis:DielectricLoss}.

The circuit of interest in this Letter is very similar to the
radio-frequency (rf) SET \cite{RF-SET} except that the SET is replaced
by a single-electron box (SEB) (see Fig.~\ref{fig:intro}a). The SEB
consists of an aluminum island connected by a tunnel junction to a
charge reservoir and three different capacitive gates. There is an rf
gate which couples the SEB to an on-chip lumped-element resonator. The
SEB also has a dc gate to adjust the static potential of the island
and a microwave (mw) gate used for spectroscopic characterization of
the SEB. We perform rf reflectometry measurements by sending a
continuous rf signal to the device. On resonance, the rf signal will
excite the resonator, which in turn will drive the potential of the
SEB island around the working point set by the dc gate. Depending on
the dissipation in the resonator caused by the SEB, the magnitude of
the reflected signal will change. We measure both the magnitude and
phase of the reflected signal to characterize this dissipation.

To explain the observed dissipation, we can consider the two energy
levels, $E_0$ and $E_1$, in Fig.~\ref{fig:intro}d. The energy of
each level can be controlled by the parameter $n_g$, the normalized
gate voltage in units of induced electrons. At the degeneracy point,
$n_g=0.5$, the two levels cross. Let us now assume that the system is
biased at a point $n_g^0$ and driven around this point by a fast rf
drive of amplitude $\delta n_g$. We imagine following the system
through one cycle of the rf drive (see Fig.~\ref{fig:intro}d). In the
first half cycle, we start on the left (right) side of the degeneracy
point in the ground state. Moving across the degeneracy point at a
high rate the system stays in the same state, which becomes the
excited state. As we move away from the degeneracy point, the
tunneling rate, $\Gamma_{+(-)}$, for electrons on to (off from) the
island, increases (see Fig.~\ref{fig:intro}e) until a tunneling event
occurs. The energy that had been put into the system is now dissipated
and we have to start over again in the second half cycle.
In contrast, when we are far from degeneracy the energy put into the
system during the first half cycle is given back during the second.

The electrostatic energy of the SEB is $E=E_C(n-n_g)^2$ where $n$ is
the number of electrons added to the island from its neutral state and
$n_g$ is the charge induced on the gate capacitance by the external
voltage. Here $E_C = e^2/2C_{\Sigma}$ is the charging energy and
$C_{\Sigma}$ is the total capacitance of the SEB island. For low
enough temperature, $k_BT \ll E_C$, and a gate charge ranging between
0 and 1, only two charge states ($E_0$ and $E_1$) need to be
considered. The dynamics of the driven SEB is then governed by the
following master equation (ME)
\begin{subequations} \label{eqn:ME}
 \begin{align}
   \dot{P}_0 &= \Gamma_{-} P_1 - \Gamma_{+} P_0 \\
   \dot{P}_1 &= \Gamma_{+} P_0 - \Gamma_{-} P_1
 \end{align}
\end{subequations}
where $P_0$ and $P_1$ are the probabilities of being in the two charge
states and the tunneling rates are given by the orthodox theory for
single-electron tunneling \cite{MPinS:Orthodox}
\begin{gather}\label{eqn:rates}
 \Gamma_{\pm} = \frac{\mp\Delta E/h}{1-e^{\frac{\pm\Delta E}{k_B T}}}
 \frac{R_K}{R_T}
\end{gather}
and where $\Delta E = E_1 - E_0 = E_C \left( 1 - 2 n_g(t) \right)$,
$n_g(t) = n_g^0 + \delta n_g \sin (\omega_0 t)$ and $\omega_0=2\pi
f_0$ is the angular frequency of the rf drive. Here $n_g^0 = C_g^{dc}
V_g/e$ is the dc gate charge, $R_K=h/e^2$ is the resistance quantum
and $R_T$ the tunneling resistance of the junction in the SEB. The
normalized rf amplitude is $\delta n_g = C_g^{rf} V_g^{rf}/e$, where
$V_g^{rf}$ is the voltage amplitude inside the resonator.
In order to solve the ME, we expand the tunneling rates
(\ref{eqn:rates}) around the working point, $n_g^0$, to first order in
$\delta n_g$ and insert them into the ME. We then linearize the
equation to get a first order linear differential equation for
$P_1$ (or $P_0$) that can be solved analytically to get $P_0(t)$ and
$P_1(t)$. The average power dissipation, \textit{i.e.} the energy
transfered to the bath per cycle of the rf drive, $T=1/f_0$, can then
be calculated as
\begin{gather} \label{eqn:Psis}
 P_{Sis} = \frac{1}{T} \int_0^T dt \left[ P_1\Gamma_{-}\Delta E -
 P_0\Gamma_{+}\Delta E \right]
\end{gather}
For a resistor, $R$, driven by an AC-voltage, $\delta V$, the average
power dissipation is $P=\delta V^2/2R$. By comparing these two values
we can define the Sisyphus resistance of the SEB,
\begin{gather} \label{eqn:sisyphus}
 R_{Sis} = 2 R_T \left(\frac{C_{\Sigma}}{C_g^{rf}}\right)^2 \frac{k_B
 T}{\Delta E^0} \sinh \left( \frac{\Delta E^0}{k_B T} \right) 
 \left( 1 +  \frac{\gamma^2}{\omega_0^2} \right)
\end{gather}
where $\Delta E^0 = E_C \left( 1 - 2 n_g^0 \right)$ is the energy
splitting at the working point and $\gamma = \Gamma_{+} +
\Gamma_{-} = \frac{\Delta E^0}{h} \frac{R_K}{R_T} \coth \left(
 \frac{\Delta E^0}{2k_B T} \right)$ is the equilibration rate of the
system. This expression is only valid for $\delta n_g \lesssim
k_BT/E_C$ where the linear expansion of the tunneling rates is
a good approximation. In this limit of small amplitudes, the
Sisyphus resistance is independent of the actual amplitude.
In addition to the Sisyphus resistance, the impedance of the SEB can
also have a reactive component.  However, our theoretical estimates
predict that the gate dependence of this reactance is negligible in
our experiment.

The device was fabricated in a multilayer process. Starting from a
high-resistivity silicon wafer with a native oxide, the wafer was
first cleaned using rf back sputtering directly after which a 60 nm
thick layer of niobium was sputtered. To pattern the niobium, we used
an Al mask made by e-beam lithography and e-beam evaporation. The
niobium was then etched in a CF4 plasma to form the inductor and
bottom plates of the capacitors (see Fig.~\ref{fig:intro}a). The Al
mask was removed with a wet-etch based on phosphoric acid. Using
PE-CVD, we then deposited an insulating layer of 200\,nm of silicon
nitride which covers the whole wafer; connections to the niobium layer
were only made through capacitors. After using a combination of DUV
photolithography to define bonding pads along with e-beam lithography
to define the top layer of the capacitors, a 3/80/10 nm thick layer of
Ti/Au/Pd was deposited by e-beam evaporation. Finally, the layer
containing the SEB was made by e-beam lithography and two-angle shadow
evaporation of 30+50 nm of aluminum, with 6 min of oxidation at 4 mbar.

The device was cooled in a dilution refrigerator with a
base temperature of about 20 mK. For the readout, we used an Aeroflex
3020 signal generator to produce the rf signal. The signal was heavily
attenuated and filtered and was fed to the tank circuit via two
Pamtech circulators positioned at the mixing chamber. The reflected
signal was amplified by a Quinstar amplifier at 4K with a nominal
noise temperature of 1 K.  The in-phase and quadrature component of
the signal were finally measured using an Aeroflex 3030 vector
digitizer.

In order to determine various device parameters, we performed
dressed-state spectroscopy of the box in the superconducting state
\cite{Wilson:LDS}.  The quantum capacitance of the box was used for
readout \cite{Duty:QC,Johansson:QLR}.
In order to determine the charging energy, we applied microwave
frequencies of $f_{\mu} = $10-11 GHz to the mw gate while slowly
sweeping the dc-gate. Multiphoton resonances occur when $mhf_{\mu}
\approx 4E_C(1-n_g)$. From the positions of the multiphoton
resonances, we can determine the charging energy
\cite{Wilson:LDS}. For this device, $E_C/h = 15.0 \pm 0.1$ GHz which
corresponds to a total capacitance of $C_{\Sigma} =1.29$ fF for the
SEB. Using this value of $E_C$, we then extracted a value for the
Josephson coupling energy of $E_J/h = 3.6 \pm 0.2$ GHz from
conventional spectroscopy.

For the normal state measurements, we applied a parallel magnetic field
of 600 mT in order to suppress superconductivity in the Al but not in
the Nb. The resonator circuit was first characterized by measuring the
reflection coefficient, $S_{11}$, as a function of frequency. Fitting
the real and imaginary part of $S_{11}$ simultaneously, the parameters
of the resonator could be determined. From the fit we extracted $L =
322$ nH, $C_c $= 92.1\,fF, $C$=109.5\,fF and
$R_{loss}$=608\,k$\Omega$. This corresponds to a resonant
frequency of $f_0$=624.7\,MHz and a total Q-value of 97. The total
Q-value has contributions both from the coupling to the external
impedance of $R_0 = 50\,\Omega$, $Q_0 = (C+C_c)/C_c^2 \omega_0^2 R_0 =
121$, and from internal losses, $Q_R = \omega_0 R (C+C_c)$, where
$R=R_{loss} R_{Sis}/(R_{loss} + R_{Sis})$. Away from the degeneracy
point, where $R = R_{loss}$, we find $Q_R = 481$. The main
contribution to the internal loss, which we represent by $R_{loss}$,
is probably due to the top layer of the capacitors which was made out
of gold. For similar circuits with superconducting Al as the top
layer, we see much less internal loss.

\begin{figure}[tb!]
 \includegraphics[width=0.9\columnwidth]{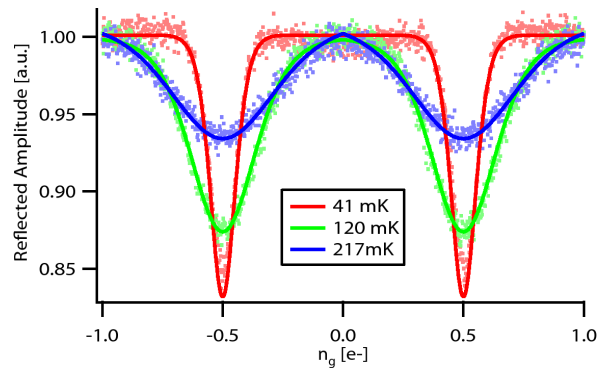}
 \caption{\label{fig:temperature}
   The reflected amplitude, for low rf amplitudes ($\delta n_g \approx
   0.04$), as a function of the the bias point, $n_g^0$, for three
   different temperatures. The theory (solid lines) are fit to the
   measured data (dots) using the analytic formula for the Sisyphus
   resistance, (\ref{eqn:sisyphus}), inserted into the expression for
   the reflection coefficient of the combined system. }
\end{figure}

Measurements of the ac-response were performed by slowly sweeping the
dc-gate and continuously monitoring $S_{11}$ at the fixed frequency
$f_0$. There was little or no change in the phase of $S_{11}$ but
there was a substantial reduction in its magnitude when the SEB was close
to its degeneracy points at $n_g = \pm 0.5$. The measurements were
done for a range of temperatures and rf amplitudes. In
Fig.~\ref{fig:temperature}, the measured response is shown for three 
different temperatures as a function of the dc-gate charge. Here we
used a low enough amplitude ($\delta n_g \approx 0.04$) that the
observed response was amplitude independent. The measured data were
then fit to theory by replacing the SEB by its Sisyphus resistance
using (\ref{eqn:sisyphus}) and calculating the reflection coefficient
for the combined system at resonance,
\begin{multline} \label{eqn:S11}
 S_{11}(\omega_0) = \frac{Z_L(\omega_0)-R_0}{Z_L(\omega_0)+R_0}
 = \frac{k\left( Q_0-Q_R \right) - i}{k\left( Q_0+Q_R \right) + i},\\
 Z_L(\omega_0) = \frac{1}{j\omega_0 Cc + \omega_0^2 C_C^2 R},
\end{multline}
where $Z_L(\omega_0)$ is the impedance on resonance of the total load, 
as pictured in Fig.~\ref{fig:intro}b, and $k = C_c/(C+C_c)$. The two fitting
parameters were $R_T$ and $C_g^{rf}$, which we were not able to
determine by independent measurements. The two higher temperature
curves were fit simultaneously, giving $R_T$=49\,k$\Omega$ and
$C_g^{rf}$=0.31\,fF. The extracted value of $R_T$ agrees with the
measured value of $E_J$ within the experimental error. The last curve
was then fit by adjusting only the temperature, yielding an electron
base temperature of $T = 41$ mK. 

\begin{figure*}[tb!]
 \includegraphics[width=\textwidth]{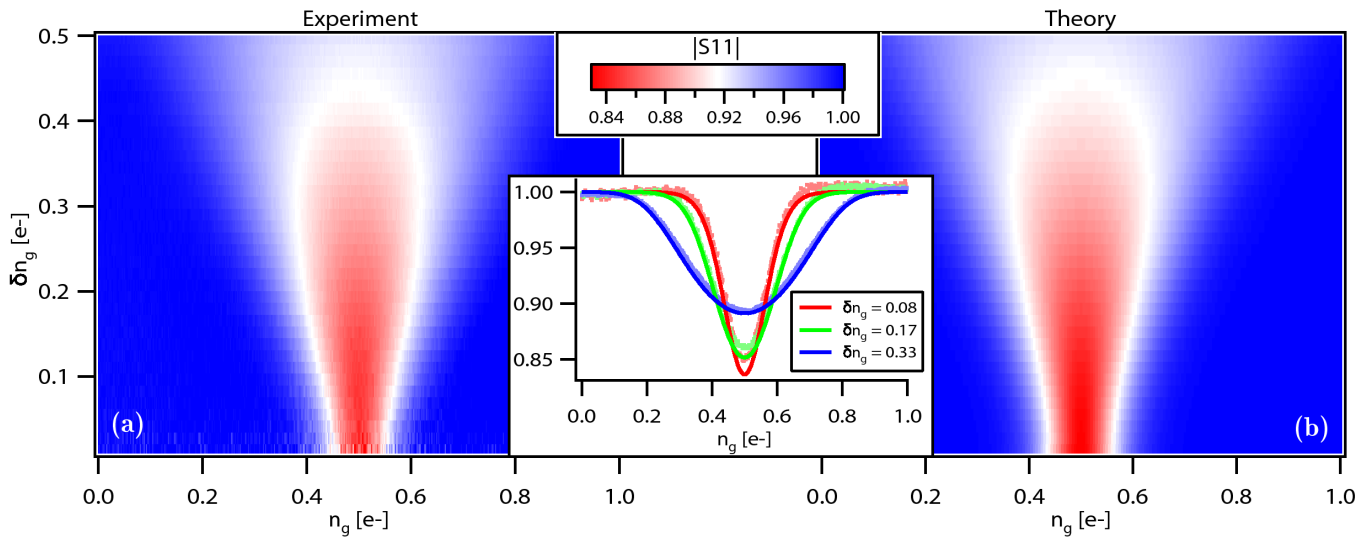}
 \caption{\label{fig:amplitude2D}
   The normalized reflected amplitude, $|S_{11}|$, measured at the
   base temperature, as a function of the bias point, $n_g^0$, and the
   rf amplitude, $\delta n_g$. In (a) are the measured data and in
   (b) is the calculated response, obtained by numerically solving the
   full ME.  The only adjustable parameter is the total rf coupling,
   which sets the y-scale. In the inset, we show three line cuts
   comparing the measured (dots) and calculated response (solid
   lines). We see a very good agreement between theory and data.}
\end{figure*}

We see that at the degeneracy point we get roughly $17\%$
absorption. This is a factor of eight more than the $2\%$ that is
expected from the naive circuit model shown in Fig.~\ref{fig:intro}c. 
If we compare the dissipated power due to the Sisyphus resistance,
$P_{Sis}$ (Eq.~(\ref{eqn:Psis})), to the dissipation, expected from
this naive circuit model, $P_{cir}$, we get that $P_{Sis}/P_{cir} =
\frac{1}{2}\left( 1 + (\omega_0 R_T C_{\Sigma})^2 \right) / \left(
  (k_B T/E_C)^2 + (\omega_0 R_T C_{\Sigma})^2 \right) \sim 8$.
A simple way to understand the large discrepancy is to consider the
voltage that develops over the tunnel junction in the two cases. In
between tunneling events, the effective resistance of the blockaded
junction is much higher than RT.  Thus, the typical voltage
developed across the blockaded junction is much larger than would
develop across a junction that is not blockaded.  The typical energy
scale of the dissipative tunneling events is therefore larger. This
leads to the curious result that, due to the Sisyphus effect, the
dissipation in the SEB is actually significantly \emph{increased} by
the Coulomb blockade, which is the opposite of what is typically found
in single-electron devices. At the degeneracy point, there is a small
discrepancy between the data and theory at the lowest
temperatures. This could possibly be explained by temperature
dependent renormalization effects \cite{Rodionov:Renormalization}.

Using the parameters extracted from the measurements in
Fig.~\ref{fig:temperature}, we can then calculate the response for 
arbitrary $\delta n_g$ and $T$.  In Fig.~\ref{fig:amplitude2D}, we
show the measured response at the base temperature as a function of
$\delta n_g$ together with the calculated response, obtained by
numerically solving the full ME.  The only additional fitting
parameter is the total rf coupling, which sets the y-scale of the
data.  We see a very good agreement between experiment and theory.

Since the Sisyphus resistance has such a strong gate dependence, it
is possible to use it as the basis for a very sensitive
electrometer. If the SEB is biased on the side of the degeneracy point
where the response has the maximum slope, a small variation in the
gate charge will give a large change in the magnitude of the
reflection coefficient. We have estimated the sensitivity, $\delta Q =
e(\delta V / V)\left|\partial S_{11}/\partial n_g\right|^{-1} $
where $V $and $\delta V$  are the applied signal and noise
voltage at the input of the amplifier, respectively.  We calculate
$\delta V$ by assuming a system noise temperature of 1.5 K due to
noise added by the amplifiers and the insertion loss between the
sample and the first amplifier. We used the calculated response in
Fig.~\ref{fig:amplitude2D} to calculate the transfer coefficient,
$\delta |S_{11}|/\delta n_g$, as a function of rf amplitude. We get a
best sensitivity of 74 $\mu\text{e}/\sqrt{\text{Hz}}$. This is not to
far from the best measured sensitivity which we obtained: 86
$\mu\text{e}/\sqrt{\text{Hz}}$. 
In order to optimize the sensitivity, as a first approximation, we
maximize the transfer coefficient $\left|\partial S_{11}/\partial
 n_g\right|$ using Eq.~(\ref{eqn:S11}) and the analytical
expression (\ref{eqn:sisyphus}). If we use an applied voltage
corresponding to $\delta n_g = k_BT/E_C$ we get that
\begin{gather}
 \delta Q(n_g^0) = \frac{k C_g^{rf}\delta V}{\sqrt{2}} Q_R(Q_0+Q_R) 
 \left( \frac{\partial Q_R}{\partial x} \right)^{-1}
\end{gather}
where $R(n_g^0) = R_{loss}\|R_{Sis}(n_g^0)$ and $x(n_g^0) =
\frac{\Delta E^0(n_g^0)}{k_BT}$. Inserting the parameters for this
sample results in a best sensitivity of $117
\mu\text{e}/\sqrt{\text{Hz}}$. Although there is a deviation from the
numerical calculation, which can be attributed to a too low estimate of
the optimal probe amplitude, the formula still give us valuable
information on how to improve the sensitivity. By reducing the losses
in the resonator to reach an internal Q-value of $10^4$ (which we have
obtained in similar resonator with a superconducting top layer) and by
decreasing the coupling capacitance, $C_c$, and the junction
capacitance, $C_J$, by a factor three each we estimate that the
sensitivity can be improved by an order of magnitude.

The use of an rf-SEB instead of the more common rf-SET has some
advantages. First, it is a simpler circuit, requiring just one tunnel
junction, which is beneficial in, for example, molecular electronics
where junctions are very difficult to fabricate. In fact, in
Ref.~\onlinecite{SEForceReadout} the same mechanism of dissipation
that has been discussed here was used to probe the potential inside
nanostructures. The model for the response developed in this Letter
could be used to understand and improve the response in such a system.
The single tunnel junction also makes the rf-SEB insensitive to
electrostatic discharges, since there is no dc path through the
device. This makes the rf-SEB more suitable for applications in
demanding environments, such as for scanning probes
\cite{Brenning:ScanningSET}. 

In conclusion, we have measured the dissipation in a single-electron
box driven by an rf signal. The observed dissipation is surprisingly
large and cannot be explained by a simple circuit model. We explain
this result using a master equation description of the Sisyphus
resistance. We also demonstrate that this phenomenon can be used for
electrometry.

We thank the members of the Quantum Device Physics and Applied Quantum
Physics groups for useful discussions. The samples were made at the
nanofabrication laboratory at Chalmers. The work was supported
by the Swedish VR and SSF, the Wallenberg foundation, and by the EU
under the project EuroSQIP. 

\bibliography{references}

\end{document}